\documentclass[10pt]{article}

\usepackage{amssymb,latexsym,amsmath,amsthm,verbatim}
\usepackage{graphicx,epsfig,epstopdf,amssymb,color}

\newcommand{\eps}{\varepsilon}
\newcommand{\be}{\begin{equation}}
\newcommand{\ee}{\end{equation}}
\newcommand{\bea}{\begin{eqnarray}}
\newcommand{\eea}{\end{eqnarray}}

\newcommand{\ba}{\begin{array}}
\newcommand{\ea}{\end{array}}
\newcommand{\R}{\mathbb{R}}
\newcommand{\Z}{\mathbb{Z}}

\def\QED{\mbox{}\hfill$\Box$}

\newtheorem{theorem}{Theorem}[section]
\newtheorem{lemma}[theorem]{Lemma}

\newtheorem{proposition}[theorem]{Proposition}
\newtheorem{corollary}[theorem]{Corollary}
\newtheorem{remark}[theorem]{Remark}

\begin{document}
\baselineskip=14pt
\title{A Simple Proof of the Stability of Solitary Waves in the Fermi-Pasta-Ulam model
near the KdV Limit}
\author{A. Hoffman and C.E. Wayne}
\maketitle
\begin{abstract}  By combining results of Mizumachi on the stability of solitons
for the Toda lattice with a simple rescaling and a careful control of the KdV limit
we give a simple proof that small amplitude, long-wavelength solitary waves in the
Fermi-Pasta-Ulam (FPU) model are linearly stable and hence by the 
results of Friesecke and Pego that they are also nonlinearly, asymptotically stable.
\end{abstract}

\section{Introduction}
In a series of four recent papers Friesecke and Pego
(\cite{friesecke:1999}, \cite{friesecke:2002}, \cite{friesecke:2004}, \cite{friesecke:2004b})  
made a detailed study of the
existence and stability of solitary wave solutions of the Fermi-Pasta-Ulam (FPU)
system:
\begin{equation}\label{eq:FPU-q}
\ddot{q}_j = V'(q_{j+1} - q_j) - V'(q_j - q_{j-1}) \qquad j \in \Z
\end{equation}
which models an infinite chain of anharmonic oscillators with nearest-neighbor interaction potential $V$.  
If we make the change of variables $r_j = q_{j+1} - q_j$ and $p_j = \dot{q}_j$, the state variable $u = (r,p)$ satisfies a system of first order Hamiltonian ODEs, 
\begin{equation}\label{eq:FPU}
u_t = JH'(u)
\end{equation}
where the Hamiltonian $H$ is given by 
\begin{equation}\label{eq:Hamiltonian}
H(r,p) = \sum_{k \in \Z} \frac{1}{2} p_k^2 + V(r_k)
\end{equation}
The symplectic operator $J$ is given by $J = \left( \ba{cc} 0 & S-1 \\ 1 - S^{-1} & 0 \ea \right)$ where $S$ is the left shift on bi-infinite sequences, i.e. $(Sx)_n := x_{n+1}$.  The problem is well posed in each $\ell^p$ space, but for concreteness and simplicity we work in $\ell^2$.  Throughout the paper we shall assume that the interaction potential $V$ satisfies the following
\begin{equation}
V \in C^4; \qquad V(0) = V'(0) = 0; \qquad V''(0) > 0; \qquad  V'''(0) \ne 0 
\label{eq:Vhyp}
\end{equation}

In the first paper in the series, \cite{friesecke:1999},  Friesecke and Pego prove that the system
\eqref{eq:FPU} has a family of solitary wave solutions which in the small amplitude, 
long-wavelength limit have a profile close to that of the KdV soliton.  More precisely
they show:

\begin{theorem}[Theorem 1.1 (b) from \cite{friesecke:1999}, restated]\label{th:FP}
Assume that $V$ satisfies $\eqref{eq:Vhyp}$ and that $c > \sqrt{V''(0)}$ is sufficiently close to $\sqrt{V''(0)}$.  Then there exists a solution to the wave profile equation for FPU:
\[ cr_c''(x) = (S + S^{-1} - 2I)V'(r_c) \qquad  \; r_c(\pm \infty) = 0\]
which in addition satisfies
\begin{equation}
\left\| \frac{1}{\eps^2} r_c\left(\frac{\cdot}{\eps}\right) - \phi_1 \right\|_{H^2} \le C\eps^2 
\label{eq:KdVclose}
\end{equation} where $\phi_1(x) := \frac{V''(0)}{V'''(0)}\left(\frac{1}{2} \mathrm{sech} (\frac{1}{2} x)\right)^2$ is the KdV soliton and $\eps := 24\sqrt{\frac{c}{V''(0)} -1}$.
\end{theorem}

\begin{remark} In fact, in \cite{friesecke:1999} it is only proven that the traveling wave
profile is close to the (rescaled) KdV soliton in the $H^1$ norm.  The strengthening of
the estimate to hold in the $H^2$ norm was done in \cite{hoffman:2008}.
\end{remark}

\begin{remark} The first general results about the existence of traveling waves in these
general FPU type systems were obtained by Friesecke and Wattis \cite{friesecke:1994} via
variational methods.
\end{remark}

In the second paper in this series, \cite{friesecke:2002}, the authors use
the method of modulation equations to prove that if the solitary waves are linearly
stable, they are nonlinearly stable as well.  More precisely, suppose that the following
linear stability condition is satisfied: \hfill \break
Define 
\begin{equation}
\omega((r,p),(\rho,\pi)) := \sum_{j \in \Z} \left(\sum_{k = -\infty}^0 p_{k+j}\rho_j +   \sum_{k = -\infty}^{-1} r_{k+j} \pi_j \right)
\label{eq:omega_def}
\end{equation}
and define
\begin{equation}
\| x \|^2_a := \sum_{k \in \Z} e^{2aj}x_j^2. 
\label{eq:anorm_def}
\end{equation}
{Hypothesis L:} {\it There are positive constants $K$ and $\beta'$ and $c_0 > \sqrt{V^{\prime \prime}(0)}$  such that 
whenever  $\sqrt{V^{\prime \prime}(0)} < c_* < c_0$, and 
 $w$ is a solution of the linear equation
\begin{equation}
\partial_t w = JH''(u_{c_*}(\cdot - c_*t))w 
\label{eq:L}
\end{equation}
with $\|w(t_0)\|_a < \infty$ and such that 
\begin{equation}
\omega(\partial_z u_{c_*}(z)|_{z = \cdot - c_*t},w(t_0)) = \omega(\partial_c u_c(z)|_{c=c_*, z = \cdot - c_*t},w(t_0)) = 0
\label{eq:sym_orth}
\end{equation}
holds, then the estimate
\begin{equation}
e^{-c_*t} \|w(t)\|_a \le Ke^{-\beta'(t-s)} e^{-ac_*s} \|w(s)\|_a
\label{eq:Lest}
\end{equation}
holds for all $t \ge s \ge t_0$.}

\begin{theorem}[Theorem 1.1 from \cite{friesecke:2002}, restated]
Assume that $V$ satisfies $\eqref{eq:Vhyp}$.  
Assume that the semigroup generated by the FPU model linearized about
its solitary wave satisfies Hypothesis L.
Then the solitary wave $u_c$ is stable in the nonlinear system 
$\eqref{eq:FPU}$ in the following sense:
Let $\beta \in (0,\beta')$.  Then there are positive constants 
$C_0$ and $\delta_0$ such that if for some $\delta \le \delta_0$ and $\gamma_* \in \R$ the initial data satisfy
\[ \|u_0 - u_{c_*}(\cdot - \gamma_*) \| \le \sqrt{\delta} \qquad \mbox{ and } \qquad \| e^{a(\cdot - \gamma_*)}(u_0 - u_{c_*}(\cdot - \gamma_*)) \| < \delta \]
then there is an unique asmyptotic wave speed $c_\infty$ and phase $\gamma_\infty$ such that 
\[ |c_\infty - c_*| + |\gamma_\infty - \gamma_*| \le C_0\delta \]
and
\[ \|u(t,\cdot) - u_{c_\infty}(\cdot - c_\infty t - \gamma_\infty) \| \le C_0 \sqrt{\delta} \qquad t > 0,\]
and in addition
\[ 
\|e^{a(\cdot - c_\infty t - \gamma_\infty)} \left( u(t,\cdot) - u_{c_\infty}(\cdot - c_\infty t - \gamma_\infty) \right) \| \le C_0 \delta e^{-\beta t} 
\]

\end{theorem}

The last two papers in the series, \cite{friesecke:2004}, \cite{friesecke:2004b},
are devoted to verifying that the linear estimate \eqref{eq:Lest} holds for the solitary
waves constructed in \cite{friesecke:1999}.

In \cite{friesecke:2004}, Friesecke and Pego construct a type of Floquet theory
to prove estimates like \eqref{eq:Lest}.  The reason that one needs such an approach
is that because of the discreteness introduced by the lattice, the linearized equation
\eqref{eq:L} is not autonomous in a frame of reference moving with the traveling waves, but
only periodic (with a spatial translation.)  Finally, in paper \cite{friesecke:2004b}, Friesecke and Pego
verify that the solitary waves constructed in \cite{friesecke:1999} satisfy the hypotheses
of their Floquet theory and hence, by the results of \cite{friesecke:2002} are asymptotically
stable.  This last step involves, among other things, the fact that these solitary waves are
well approximated by the KdV soliton and the fact that the linearization of the
KdV equation about its solitary wave is well understood.

In this note, we give a simple alternative proof of the estimate \eqref{eq:Lest} which avoids
the use of the Floquet theory and spectral analysis of \cite{friesecke:2004} and \cite{friesecke:2004b}.

Our proof combines three observations:
\begin{enumerate}
\item The linear stability of the soliton for the Toda lattice, established by Mizumachi
and Pego by the construction of an explicit B\"acklund transformation in \cite{mizumachi:2007}.
\item A transformation of the original FPU equation \eqref{eq:FPU} into a form
in which we can prove that its solitary wave solution is close to that of the Toda lattice.
\item A careful control of the way in which various quantities depend on the small parameter
$\epsilon$.
\end{enumerate}

We note that the last two points were originally developed in our study of counter-propagating
2-soliton solutions of the FPU model \cite{hoffman:2008}.

We now explain these three points in more detail.  Recall first that the Toda lattice
is the special case of the FPU model with potential function
\begin{equation}\label{eq:Toda_potential}
\tilde{V}(x) = (e^{-x} +x-1)\ .
\end{equation}
(Throughout this paper, quantities with tildes will refer to the Toda model.)  Note that
$\tilde{V}$ satisfies the hypotheses of \cite{friesecke:1999} so the results of that
paper imply that the Toda model has a family of solitary waves close to those of the KdV
equation.  Of course in the case of the Toda model these
solutions were explicitly constructed by Toda in the 1960's and indeed the Toda model is
one of the classic examples of a completely integrable, infinite-dimensional Hamiltonian
system.  The stability of the Toda soliton can also be analyzed in a very direct fashion.
By constructing a B\"acklund transformation which conjugates the linearization of the
Toda model about its soliton to the linearization of the Toda lattice about zero,
Mizumachi and Pego proved that the linearized Toda equation satisfies Hypothesis L and 
hence, by the results of \cite{friesecke:2002}, that the Toda soliton is asymptotically stable.  
In \cite{hoffman:2008} we extended this result by showing that the constant $K$ in $\eqref{eq:Lest}$ can be chosen uniformly in $c$.

\begin{remark}  Although we will be most interested in these results in the long-wavelength,
small amplitude regime studied by Friesecke and Pego, in fact the results of \cite{mizumachi:2007}
apply to Toda lattice solitons of arbitrary size.
\end{remark}

In order to extend the estimate on the linear decay from the semigroup of the linearized Toda
equation to the linearization of the general FPU model satisfying hypothesis {\bf (H1)}, we first
make use of the following simple:
\begin{lemma} \label{lemma:rescaling} Without loss of generality we may assume that the 
potential energy function $V$ in \eqref{eq:FPU-q} (or equivalently \eqref{eq:Hamiltonian})
satisfies
$$
V^{\prime \prime}(0) = 1\ ,\ \  V^{\prime \prime \prime}(0) = -1\ .
$$
\end{lemma}

\proof To see this simply note that if $q(j,t)$ is a solution of \eqref{eq:FPU-q}
with a potential energy function $V$ that satisfies hypothesis {\bf (H1)}, then
$V$ can be written as $V(x) =   \frac{1}{2} \alpha x^2 + \frac{1}{6} x^3 + {\cal O}(x^4)$
for some $a>0$ and $b \ne 0$.   If we now define
$\hat{q}(j.t) = \alpha q(j,\beta t)$, then $\hat{q}$ solves the FPU equations with potential
function $\hat{V}(x) = \frac{1}{2} \alpha x^2 + \frac{1}{6} b (\beta^2/\alpha) x^3 + {\cal O}(x^4)$,
so choosing $\alpha$ and $\beta$ appropriately insures that the lemma holds.
\qed

From now on we will assume that the potential function $V$ satisfies this normalization.
Note that with this normalization the potential $V$ in \eqref{eq:FPU-q} differs from the
Toda potential only by terms of ${\cal O}(x^4)$ or higher.  With this observation the
existence results of \cite{friesecke:1999} imply:
\begin{proposition}\label{prop:Toda-difference}  Let $u_c$ by the profile of the solitary
wave of the FPU model \eqref{eq:FPU-q} with speed $c$, and let $\tilde{u}_c$ be the profile for the
special case of the Toda potential.  Define $\xi_1 = \partial_{z}u_c(z)$ and $\xi_2
= \partial_c u_c(z)$ and let $\tilde{\xi}_{1,2}$ be the corresponding quantities for the Toda
lattice.  Then there exists $\epsilon_0, C > 0$ such that for $0<\epsilon<\epsilon_0$,
one has the estimates
\begin{eqnarray}
\| u_c -  \tilde{u_c} \|_{\ell^2_a} \le C \epsilon^{7/2} && \| u_c -  \tilde{u_c} \|_{\ell^{\infty}} \le C \epsilon^{4} \\
\| \xi_1 -  \tilde{\xi}_1 \|_{\ell^2_a} \le C \epsilon^{9/2} && \| \xi_1 -  \tilde{\xi}_1 \|_{\ell^{\infty}} \le C \epsilon^{5} \\
\| \xi_2 -  \tilde{\xi}_2  \|_{\ell^2_a} \le C \epsilon^{3/2} && \| \xi_2 -  \tilde{\xi}_2 \|_{\ell^{\infty}} \le C \epsilon^{2} 
\end{eqnarray}
\end{proposition}
\begin{remark} The half-powers of $\epsilon$ that occur in the estimates of the $\ell^2_a$
norms are a consequence of the scaling of the functions $u_c$.
\end{remark}
\proof The inequality $ \| u_c -  \tilde{u_c} \|_{\ell^{\infty}} \le C \epsilon^{4}$ follows immediately from $\eqref{eq:KdVclose}$ because to leading $(\eps^2)$ order both $u_c$ and $\tilde{u}_c$ agree with the KdV soliton.  The estimate on $\| u_c -  \tilde{u_c} \|_{\ell^2_a}$ then follows from this estimate
because of the prior remark about the scaling of the $\ell^2_a$ norms.  The estimates on 
$ \xi_1 -  \tilde{\xi}_1$ then follow from these two since Theorem \ref{th:FP} shows that
a derivative of the solitary wave profile with respect to the spatial variable gains exactly
one power of $\epsilon$.  The estimates for $ \xi_2 -  \tilde{\xi}_2 $ follow in a similar fashion.
For more details see \cite{hoffman:2008}.  
\QED

With these estimates in hand we consider the semi-group generated by
\begin{equation}\label{eq:linear_semigroup}
\partial_t v = J H^{\prime \prime}(u_c) v = J \tilde{H}^{\prime \prime}(\tilde{u}_c) v
+ J \left(H^{\prime \prime}(u_c) - \tilde{H}^{\prime \prime}(\tilde{u}_c) \right) v\ .
\end{equation}
The idea is now to treat the term $J \left(H^{\prime \prime}(u_c) - \tilde{H}^{\prime \prime}(\tilde{u}_c) \right) v$ as a perturbation of the Toda semi-group.  Recalling that $H$ and $\tilde{H}$ differ
only at quartic order and that $u_c$ and $\tilde{u}_c$ are both of order ${\cal O}(\epsilon^2)$
and differ only by terms of ${\cal O}(\epsilon^4)$ we have 
\begin{eqnarray}
\| J \left(H^{\prime \prime}(u_c) - \tilde{H}^{\prime \prime}(\tilde{u}_c) \right) v \|_{\ell^2_a} 
&\le & C\left( \| u_c - \tilde{u}_c \|_{\ell^{\infty}} + ( \| u_c \|_{\ell^{\infty}} + \| \tilde{u}_c \|_{\ell^{\infty}})^2 \right) \| v \|_{\ell^2_a} \\  \nonumber & \le & C \epsilon^4 \| v \|_{\ell^2_a}\ .
\end{eqnarray}

The other fact we must deal with is that $v \in E^s = \{ v ~|~ \omega(\xi_1,v) = \omega(\xi_2,v) =0 \}$,
while our decay estimates on the Toda semigroup hold only if the semigroup acts
on vectors $\tilde{v} \in \tilde{E}^s = \{ \tilde{v} ~|~ \omega(\tilde{\xi}_1,\tilde{v}) = 
\omega(\tilde{\xi}_2,\tilde{v}) =0 \}$.  To cope with this difference we define the projection
operator
\begin{equation}
Q v = v- \left( \frac{ \omega(\tilde{\xi}_2,\tilde{\xi}_1) \omega(\tilde{\xi}_2, v) + 
 \omega(\tilde{\xi}_2,\tilde{\xi}_2) \omega(\tilde{\xi}_1, v)}{ \omega(\tilde{\xi}_2,\tilde{\xi}_1)^2 }\right) 
 \tilde{\xi}_1 - \frac{  \omega(\tilde{\xi}_1, v) }{\omega(\tilde{\xi}_2,\tilde{\xi}_1) } \tilde{\xi}_2\ ,
\end{equation}
which maps $\ell^2_a$ to $\tilde{E}_s$.  Using the estimates in Proposition 
\ref{prop:Toda-difference} , we have:
\begin{proposition}\label{prop:projection}[Lemma 4.4 in \cite{hoffman:2008}, simplified]  There exists $\epsilon_0, C  > 0$ such that
if $ 0 < \epsilon < \epsilon_0$, then
$$
\| Q v \|_{\ell^2_a} \le C \| v  \|_{\ell^2_a}\ .
$$
Furthermore, if $v \in E^s$, then
$$
\| ( I - Q ) v \|_{\ell^2_a} \le C \epsilon^{3/2} \| v \|_{\ell^2_a}\ .
$$
\end{proposition}
\proof  The key factor in the proof of this proposition is that the
estimates of Theorem \ref{th:FP} make it possible to evaluate the
leading order in $\epsilon$ behavior of the symplectic products
$\omega({\xi}_j,\xi_k)$ and $\omega(\tilde{\xi}_j,\tilde{\xi}_k)$.
This was first used in \cite{friesecke:1999} and  was utilized
repeatedly in \cite{hoffman:2008}.  For instance, 
one has $\omega(\xi_1,\xi_1)=0$, while $\omega(\xi_2,\xi_2)
= c_{22} \epsilon^{-2} + {\cal O}(\epsilon^{-1})$ and the
cross term $\omega(\xi_1,\xi_2)= c_{12} \epsilon + {\cal O}(\epsilon^2)$
with the constants $c_{12}$ and $c_{22}$ both non-zero.
Similarly, the leading order behavior in $\epsilon$ of the
norms of $\xi_1$ and $\xi_2$ can be computed by relating them
to the derivitives of the profile of the KdV soliton using
the estimates of Theorem \ref{th:FP}.
The same estimates also hold for the symplectic inner products
of $\tilde{\xi}_{1,2}$ and with these estimates the first
bound in the proposition follows immediately.

The second estimate follows by rewriting the projection operator
as 
\begin{eqnarray}
\omega(\tilde{\xi}_2,\tilde{\xi}_1) ( I - Q ) v & = & \left( \frac{ \omega(\tilde{\xi}_2,\tilde{\xi}_1) \omega(\tilde{\xi}_2, v) + 
 \omega(\tilde{\xi}_2,\tilde{\xi}_2) \omega(\tilde{\xi}_1, v)}{ \omega(\tilde{\xi}_2,\tilde{\xi}_1) }\right) 
 \tilde{\xi}_1 - \omega(\tilde{\xi}_1, v) \tilde{\xi}_2  \\ \nonumber
 & = & \left( \frac{ \omega(\tilde{\xi}_2,\tilde{\xi}_1) \omega(\tilde{\xi}_2 -\xi_2, v) + 
 \omega(\tilde{\xi}_2,\tilde{\xi}_2) \omega(\tilde{\xi}_1, v)}{ \omega(\tilde{\xi}_2,\tilde{\xi}_1) }\right) 
 \tilde{\xi}_1 - \omega(\tilde{\xi}_1-\xi_1, v)  \tilde{\xi}_2\ ,
\end{eqnarray}
where the last step used the fact that since $v \in E^s$, $\omega(\xi_1,v)
=\omega(\xi_2,v) =0$.  But now note that each term on the right hand side contains
a factor of either $\tilde{\xi}_1 - \xi_1$ or $\tilde{\xi}_2 - \xi_2$ and these are small due to
the estimates in Proposition \ref{prop:Toda-difference} .  With the aid of these estimates
the second estimate in the proposition follows in a straightforward fashion.
For more details see the proof of Lemma 4.4 in \cite{hoffman:2008}.
\QED

\begin{corollary}\label{corr:projection} There exists $\epsilon_0  > 0$ such that
if $ 0 < \epsilon < \epsilon_0$ and $v \in E^s$, then
$$
\| v  \|_{\ell^2_a}  \le 2 \| Qv  \|_{\ell^2_a} \ .
$$
\end{corollary}

Now write $v(t)$, the solution of \eqref{eq:linear_semigroup} as
\begin{equation}
v(t) = \tilde{S}(t,0) v(0) + \int_0^t \tilde{S}(t,s)  J \left(H^{\prime \prime}(u_c) - 
\tilde{H}^{\prime \prime}(\tilde{u}_c) \right) v(s) ds\ ,
\end{equation}
where $\tilde{S}$ is the evolution semigroup generated by the linearized Toda system.  Then
\begin{equation}
Qv(t) = \tilde{S}(t,0) Q v(0) + \int_0^t \tilde{S}(t,s) Q\left( J \left(H^{\prime \prime}(u_c) - 
\tilde{H}^{\prime \prime}(\tilde{u}_c) \right) \right) v(s) ds\ ,
\end{equation}
Taking the norm of both sides and using the estimates on the Toda semigroup $\eqref{eq:Lest}$ we find:
\begin{eqnarray}
\| Q v  \|_{\ell^2_a} &\le&K e^{-b \epsilon^3 t} \| v(0)  \|_{\ell^2_a} +
K \int_0^t e^{-b \epsilon^3 (t-s)} \|  Q\left( J \left(H^{\prime \prime}(u_c) - 
\tilde{H}^{\prime \prime}(\tilde{u}_c) \right) \right) v(s) \|_{\ell^2_a}  ds \\
&\le& K   e^{-b \epsilon^3 t} \| v(0)  \|_{\ell^2_a} +
K_2 \epsilon^4  \int_0^t e^{-b \epsilon^3 (t-s)} \| v(s) \|_{\ell^2_a}  ds \\ 
&\le& K  e^{-b \epsilon^3 t} \| v(0)  \|_{\ell^2_a} +
K_3 \epsilon^4  \int_0^t e^{-b \epsilon^3 (t-s)} \| Q v(s) \|_{\ell^2_a}  ds 
\end{eqnarray}
Note that the constants $K$, $K_2$, and $K_3$ are all independent of $\epsilon$.
Now setting $\phi(t) = \sup_{0 \le \tau \le t} e^{b' \epsilon^3 \tau} \| Q v(\tau)  \|_{\ell^2_a}$
for some $ 0 < b' < b$ and taking the supremum in the above equation we have
\begin{eqnarray}
\phi(t) &\le &  K  \| v(0)  \|_{\ell^2_a} 
+ K_3 \epsilon^4 e^{-(b-b')  \epsilon^3 t} \int_0^t e^{-(b-b')  \epsilon^3 s} \phi(s) ds \\
&\le &  K \| v(0)  \|_{\ell^2_a} 
+ K_3 \epsilon \phi(t)\ .
\end{eqnarray}

Thus, if $\epsilon$ is sufficiently small, we conclude that $\phi(t)$ is uniformly bounded for all $t$ and hence:
\begin{proposition} \label{prop:linear}
There exists $K' > 0$, independent of $\epsilon$, and
$\epsilon_0>0$ such that if $0<\epsilon<\epsilon_0$ and $v(t) \in E^s$ is a solution of
\eqref{eq:linear_semigroup} then
\[
\| v  \|_{\ell^2_a} \le K' e^{-b' \epsilon^3 t} \| v(0)  \|_{\ell^2_a}\ .
\]
\end{proposition}

This verifies that the linearized FPU semigroup satisfies
Hypothesis L and hence by the results of \cite{friesecke:2002} that the FPU solitary wave is
asymptotically stable and the small amplitude, long-wavelength regime.

\begin{remark}  We note that in the proof of Proposition \ref{prop:linear} it is important to
carefully control the dependence of the semi-group on the parameter $\epsilon$.
It is not surprising that a perturbation arguments permits one to extend the results of 
Mizumachi and Pego to solitary waves in small perturbations of the Toda model.
Indeed, this was already noted in \cite{mizumachi:2007} .
What we do find noteworthy is that this simple argument can cover all
FPU solitary waves in the KdV regime, and this requires the detailed study of
the small $\epsilon$ asymptotics  contained in Propositions \ref{prop:Toda-difference} 
and \ref{prop:projection}.
\end{remark}

{\bf Acknowledgements:} This work was funded in part by the National Science 
Foundation under grants DMS-0603589 and DMS-0405724.
Any findings, conclusions, opinions, or recommendations are those of the authors,
and do not necessarily reflect the views of the NSF.

\bibliography{FPU_reference}

\begin{thebibliography}{1}

\bibitem{friesecke:1999}
G.~Friesecke and R.~L. Pego.
\newblock Solitary waves on {FPU} lattices. {I}. {Q}ualitative properties,
  renormalization and continuum limit.
\newblock {\em Nonlinearity}, 12(6):1601--1627, 1999.

\bibitem{friesecke:2002}
G.~Friesecke and R.~L. Pego.
\newblock Solitary waves on {FPU} lattices. {II}. {L}inear implies nonlinear
  stability.
\newblock {\em Nonlinearity}, 15(4):1343--1359, 2002.

\bibitem{friesecke:2004}
G.~Friesecke and R.~L. Pego.
\newblock Solitary waves on {F}ermi-{P}asta-{U}lam lattices. {III}.
  {H}owland-type {F}loquet theory.
\newblock {\em Nonlinearity}, 17(1):207--227, 2004.

\bibitem{friesecke:2004b}
G.~Friesecke and R.~L. Pego.
\newblock Solitary waves on {F}ermi-{P}asta-{U}lam lattices. {IV}. {P}roof of
  stability at low energy.
\newblock {\em Nonlinearity}, 17(1):229--251, 2004.

\bibitem{hoffman:2008}
Aaron Hoffman and C.~Eugene Wayne.
\newblock Counterpropagating two-soliton solutions in the {F}{P}{U} lattice.
\newblock Preprint; arXiv:0806.1637, 2008.

\bibitem{friesecke:1994}
Gero Friesecke and Jonathan A.~D. Wattis.
\newblock Existence theorem for solitary waves on lattices.
\newblock {\em Comm. Math. Phys.}, 161(2):391--418, 1994.

\bibitem{mizumachi:2007}
Tetsu Mizumachi and Robert~L. Pego.
\newblock Asymptotic stability of toda lattice solitons.
\newblock Preprint, 2007.

\end{thebibliography}

\end{document}